# Self-Diagnosis, Scaffolding and Transfer in a More Conventional Introductory Physics Problem

E. Yerushalmi[1], A. Mason[2], E. Cohen[1] and C. Singh[2]

[1] Department of Science Teaching, Weizmann Institute of Science, Rehovot, Israel
[2] Department of Physics and Astronomy, University of Pittsburgh, Pittsburgh, PA 15213, USA

**Abstract.** Previously we discussed how well students in an introductory physics course diagnosed their mistakes on a quiz problem with different levels of scaffolding support. In that case, the problem they self-diagnosed was unusually difficult. We also discussed issues related to transfer, particularly the fact that the transfer problem in the midterm that corresponded to the self-diagnosed problem was a far transfer problem. Here, we discuss a related intervention in which we repeated the study methodology with the same students in the same intervention groups, using a new quiz problem which was more typical for these students and a near transfer problem. We discuss how these changes affected students' ability to self-diagnose and transfer from the self-diagnosed quiz problem to a transfer problem on the midterm exam.

## INTRODUCTION

Previously we described a self-diagnostic task in which students identify and explain their own errors with different levels of scaffolding [1]. This task involved asking students to solve a difficult, context-rich problem [4] in a quiz and then asking them to diagnose their mistakes. We considered three different levels of scaffolding during the intervention to determine what level of support helped students self-diagnose the best. We then asked in the midterm exam a problem involving similar physical principles in a somewhat similar setup to explore whether self-diagnosis helped students in different experimental groups compared to the control group. Our primary findings were that students with a higher level of scaffolding performed better on self-diagnosis, but that transfer to the midterm problem was limited overall for all intervention groups [1-2]. Considerations of performance on the quiz problem and midterm problem suggested that the problems were very difficult for students; not a single student got full credit on the quiz problem and few were able to answer the corresponding midterm problem correctly.

While the quiz and midterm problems involved somewhat similar setups [3], they were not completely isomorphic. Both problems involved determining the forces on a known mass at the maximum/minimum point on a circular path with known radius, given information on the height and velocity in the launch point. They required recognition of similar target variables (normal or tension force) and intermediate variables (velocity and centripetal acceleration at the point of interest) and to implement conservation of mechanical energy and Newton's second law in a non-equilibrium situation involving centripetal acceleration. Yet, the problems differed in context as well in that the midterm problem included an additional initial part to the quiz problem that asked students to determine the point of interest before solving the rest of the problem (i.e., at what point in a pendulum's swinging motion is the tension in the rope the greatest?). Approximately 15% of students incorrectly believed that tension force was maximum in the rope at the highest point and therefore neither invoked the conservation of mechanical energy nor centripetal acceleration which is zero at the highest point where the pendulum is momentarily at rest. These students could not transfer from self diagnosis of the quiz problem to the midterm [5].

We decided to conduct another experiment using the same levels of scaffolding as described in the previous papers [1-3]. But in this study, another set of problems were chosen for self-diagnosis in quiz and transfer (in a different midterm) with the following criteria: a) while this second quiz problem that students self-diagnosed was also a context-rich problem with two parts, the two principles used were both more familiar and easier to apply for introductory students and b) the paired midterm problem does not require as far of a transfer from the self-diagnosed quiz problem. The goal of this paper is to examine students' ability to self-diagnose this problem with different scaffolds and their ability to transfer to the near transfer midterm problem. We again focus on how well students review the solution that they have composed themselves in order to improve it or learn from it [1-3]. We will consider both the students' knowledge content and their approach to problem-solving, as self-diagnosis may pertain to arriving at a solution as well as more general learning goals [6].

The process of "self-diagnosis" task is the same as in the previous study [1-3]. Students are required to present a diagnosis (namely, identify where they went wrong, and explain the nature of the mistakes) as part of the activity of reviewing their quiz solutions. In the present paper we consider the same research questions posed in the previous papers but in the context of a more typical quiz problem and a closer transfer task. We adapt the scoring rubric developed for the previous study [1]. This rubric allowed us to score both the solution and the self diagnosis, and to differentiate between deficiencies in the "physics" as well as the "presentation" of the solution. In a companion paper [7], we will compare the findings regarding both problem pairs.

**TABLE 1.** Distribution of groups for self-diagnosis tasks.

| | Self-diagnosis tasks | | |
|---|---|---|---|
| *Group A/A'* | *Groups B/B'* | *Group C* | *Group D* |
| control | Instructor outline, diagnosis rubric | Worked out example | Minimal guidance: notes + text books |
| ~100 students *3 sections* | 60 students *2 sections* | 26 students *1 section* | 23 students *1 section* |

## EXPERIMENTAL SETUP & SAMPLE

The study involved an introductory algebra based course for pre-meds (N~200), with one instructor and two teaching assistants (TAs). The TA classrooms were distributed into control groups and three self-diagnosis treatments groups each of which carried out self-diagnosis task with different scaffolds (see Table 1). In all treatment groups, students first solved a quiz problem (quiz 7), and in the next training session they were asked to circle mistakes in their photocopied solutions and explain what they did wrong.

The groups' initial attempts on the quiz are shown in Table 2. Note that all students made mistakes even on an easier problem and therefore performed self-diagnosis. However, average quiz scores were approximately 10 points higher than scores for the quiz reported in earlier study [1-3]).

**TABLE 2.** Grades for quiz 7.

## The Quiz and Midterm Problems

The problem used for quiz 7 shown was adapted from the on-line archive of UMN PER group [8]. To solve this problem, conservation of momentum must first be applied for the inelastic collision of the "friend" and the skateboard so that the velocity of the two moving together may be found. Then conservation of energy should be used to determine how far up the slope the skateboarding friend will go.

> You are helping a friend prepare for the next skate board exhibition. Your friend who weighs 60 kg will take a running start and then jump with a speed of 1.5 m/s onto a heavy duty 5 kg stationary skateboard. Your friend and the skateboard will then glide together in a straight line along a short, level section of track, then up a sloped concrete incline plane. Your friend wants to reach a minimum height of 3 m above the starting level before he comes to rest and starts to come back down the slope. Knowing that you have taken physics, your friend wants you to determine if the plan can be carried out or whether he will stop before reaching a 3 m height.
> Do not ignore the mass of the skateboard.

In the midterm problem, instead of a friend jumping horizontally on a skateboard, Fred Flintstone is shown jumping into his cart at an angle of 45 degrees from the downward vertical axis. The problem is well defined in comparison to the quiz context rich problem in that the students are explicitly asked to find the initial velocity of the cart with Fred in it and the maximum height that the cart will roll up a hill afterwards. This problem is similar to the quiz problem in that it employs the same principles, i.e., the conservation of momentum and the conservation of mechanical energy in a similar physical setup.

## FINDINGS

As shown in the previous study [1] we differentiated the researcher's judgment of the students' self-diagnosis into "physics" and "presentation" scores. The "physics" score has 2 sub-categories: "Invoking" and "applying". The 1st scores deficiencies in invoking the principles needed to solve the problem (e.g., in this problem, conservation of momentum and mechanical energy), in defining the system appropriately and consistently, and avoiding inappropriate principles (for which there would be no grade if not present). The "applying" subcategory evaluates the application of

| Quiz 7 | | Control | | Intervention: Outline + Rubric | | Intervention: Sample solution | Intervention: Minimal guidance |
|---|---|---|---|---|---|---|---|
| Group | | A | A' | B | B' | C | D |
| Physics | Mean | 0.47 | 0.53 | 0.45 | 0.50 | 0.40 | 0.50 |
| | Std. Err. | 0.03 | 0.04 | 0.03 | 0.04 | 0.04 | 0.04 |
| Presentation | Mean | 0.40 | 0.40 | 0.49 | 0.43 | 0.36 | 0.38 |
| | Std. Err. | 0.02 | 0.02 | 0.02 | 0.02 | 0.01 | 0.03 |

these principles. We weighed each item in these subcategories as worth 1, 2/3, 1/2, 1/3 or 0 points (corresponding respectively to marks of +, ++/-, +/-, +/-- or -). The rubric also took into account situations when a student did not invoke a principle and thus would necessarily not apply it in order to prevent "double penalizing". Presentation subcategories included: "description" of the problem (e.g. absence of free body diagram…), explicating solution "plan", and presenting an after the fact "check" of the solution.

## Self-diagnosis - Physics

The physics self-diagnosis score reflected both the *expert ideal knowledge* (correct ideas needed to solve the problem) and the *novice knowledge per se* (includes incorrect ideas student believe are needed to solve the problem as reflected in his/her solution and diagnosis) [1-3]. This approach allowed us to identify diagnosis ability of student's who do not know the correct answer but still identify something is wrong.

Tables 3 and 4 present a comparison of students' performance on self-diagnosis (SD) of physics aspects in the alternative groups. Note that even in this quiz problem, all students made mistakes, thus all of them are included in the analysis. One common tendency was to fail to invoke the momentum conservation principle altogether and implement mechanical energy conservation, assuming that the friend's velocity was the same as the velocity of the friend on the skateboard. Scores were tabulated according to the rubric outlined previously [1-3].

Tables 3 and 4 suggest that an easier problem in this quiz allowed students to make effective use of whatever resources and tools they were provided even when the scaffolding support merely allowed students to use their notes and textbook. This can be seen from the fact that even group D students did fairly reasonable self-diagnosis unlike the diagnosis performed by group D students in the previous study when the quiz problem was extremely difficult [1-3].

**TABLE 3**. Grades for students in self-diagnosis of physics mistakes.

|  | Group B | Group B' | Group C | Group D |
|---|---|---|---|---|
| Mean | 0.56 | 0.70 | 0.62 | 0.61 |
| Std. Err. | 0.056 | 0.064 | 0.06 | 0.065 |

**TABLE 4**. The p-values for table 2 grades.

| Group | B | B' | C | D |
|---|---|---|---|---|
| B |  | 0.14 | 0.45 | 0.62 |
| B' |  |  | 0.44 | 0.35 |
| C |  |  |  | 0.84 |

Table 5 presents students' performance of self-diagnosis in invoking the correct physics principles and applying them (i.e., the percentage of students who diagnosed their mistakes out of those who made mistakes in each sub category). For the self-diagnosis of the quiz problem in the previous study, it was easier to identify mistakes in invoking principles than it was to find mistakes in applying those principles, and also it was easier to invoke a correct principle than it was to apply the principle correctly. Here, the opposite is true. For all three groups, more students were unable to invoke all the correct principles than there were students who invoked principles but failed to apply them correctly. In particular, it appears that many students simply overlooked conservation of momentum and only addressed conservation of energy in quiz initially. This is consistent with prior research in which students were able to name one, but not both, of two principles involved in a problem about a ballistic pendulum [9].

**TABLE 5.** Self-diagnosis grades for physics subcategories. Explanation of symbols is as follows. "+" represents a correct diagnosis; "+/-" represents a partially correct diagnosis; and "–" represents an incorrect diagnosis or no diagnosis performed. "T" refers to the total percentages of students who had mistakes in their quiz regarding some subcategory; the students who got a subcategory correct were not included.

| Subcategory Group | Invoking | | | | Applying | | | |
|---|---|---|---|---|---|---|---|---|
| Grade | + | +/- | - | T | + | +/- | - | T |
| B | 29 | 23 | 48 | 39 | 57 | 43 | 0 | 11 |
| B' | 10 | 22 | 68 | 29 | 40 | 40 | 20 | 8 |
| C | 29 | 17 | 53 | 48 | 100 | 0 | 0 | 8 |
| D | 30 | 36 | 34 | 46 | 50 | 50 | 0 | 4 |

## Midterm - Physics

As stated earlier, the midterm problem paired with quiz problem in previous study [1-3] was a far transfer problem on which students did not perform well. Therefore, here, we selected a midterm problem which was a nearer transfer problem for the quiz problem discussed in the earlier section.

Table 6 shows the mean physics score for all groups on the midterm problem. To be able to consider the effect of the TAs on the inter-group comparison, we present analysis of each TA's groups separately. Table 7 shows ANCOVA p-value comparisons between group B and the control group and between group C and group D, respectively. One can see that group B, which was provided a rubric for presenting their self-diagnosis and solution outline, did about as well as the control group A. This finding suggests that the scaffolding provided by self-diagnosis made little difference on midterm performance. Since this midterm exam was several weeks after the self-diagnosis activities and students were provided the written solution for the quiz problem, learning outside of the in-class self-diagnosis may also be responsible for the midterm performance.

**TABLE 6**. Midterm physics grades paired with the quiz problem.

|  | First TA | | | Second TA | |
| --- | --- | --- | --- | --- | --- |
| Group | A | B | B' | C | D |
| Mean | 0.64 | 0.66 | 0.68 | 0.72 | 0.76 |
| Std. Err. | 0.025 | 0.04 | 0.048 | 0.058 | 0.043 |

**TABLE 7**. The p-values for table 3 grades.

| First TA | Group B | Group B' |
| --- | --- | --- |
| Group A | 0.73 | 0.40 |
| Group B |  | 0.67 |

| Second TA | Group D |
| --- | --- |
| Group C | 0.94 |

A surprising finding which discussed in the companion paper is that the only group for which we found positive correlation between the physics self diagnosis scores and the midterm scores was group D, which obtained the minimal scaffolding. In the companion paper, we discuss a framework for analyzing the effects of various interventions by classifying the interventions as "weak", "superficial" or "meaningful". We hypothesize the kinds of correlations (e.g., positive or negative) that would result between the quiz score and self-diagnosis score or between the self-diagnosis score and midterm score for each type of intervention and then analyze our data to understand the nature of interventions and its impact on student learning better.

Additionally, Table 6 suggests that all groups performed better on the midterm than the quiz problem and Table 7 suggests they performed roughly equivalently. This finding suggests that the transfer was sufficiently close so that students were able to take advantage of whatever scaffolding they received as well as the posted solution on course website.

## Performance on Presentation

Students in all groups performed poorly on presentation of solution in quiz and in self-diagnosing errors in problem presentation. Groups B and B' again performed slightly better in self-diagnosis presentation than groups C and D. As stated previously [3], this may be attributed to the additional scaffolding they received in order to perform the self-diagnosis (as discussed previously [3], groups B and B' were the only groups which were provided a rubric directing them explicitly to self-diagnose their presentation). However, groups B and B' fared about the same as the other groups with regard to the presentation score on the midterm (for which no rubric was provided for presenting the solution to even groups B and B'), despite a better performance on presentation self-diagnosis. The poor presentation in the midterm of all groups supports the previous assertion [3] that a sustained intervention is needed to help students develop effective problem-solving communication skills which is a habit of mind.

## SUMMARY

We described students' performance in self-diagnostic tasks that involved minimal modeling and feedback with a more conventional context-rich problem than the one featured in previous study [1-3], as well as their performance in midterm on a paired problem with a closer transfer of knowledge. The group with notes and text as scaffold (group D) performed as well as the other groups on self-diagnosis and the paired midterm problem.

In the companion paper [6], we discuss a framework for analyzing the effects of various interventions and also discuss the differences between the two studies and what we learned overall about self-diagnosis, scaffolding and transfer.

## ACKNOWLEDGMENTS

We thank ISF for 1283/05 & NSF for DUE-0442087.

## REFERENCES

1. A. Mason, E. Cohen, E. Yerushalmi, and C. Singh, AIP Conf. Proc. **1064**, 147-150, (2008).
2. E. Cohen, A. Mason, C. Singh, & E. Yerushalmi, AIP Conf. Proc. **1064**, 99-102, (2008).
3. E. Yerushalmi, A. Mason, E. Cohen, and C. Singh, AIP Conf. Proc. **1064**, 53-56, (2008).
4. P. Heller and M. Hollabaugh, *Am. J. Phys.* 60, 637, (1992).
5. C. Singh, Am. J. Phys. **77**, 73 (2009).
6. J. Larkin, J. McDermott, D. Simon, and H. Simon, *Science* **208**, 1335-1342, (1980).
7. A. Mason, E. Cohen, C. Singh, and E. Yerushalmi, 2009 PERC proceedings, AIP Conf. Proc., Melville, New York **1179**, 27-30 (2009).
8. http://groups.physics.umn.edu/physed/Research/CRP/on-lineArchive/crcecm.html#ce
9. C. Singh and D. Rosengrant, *Am. J. Phys*. 71(6), 607-617, (2003).